\documentclass[aps,prapplied,twocolumn,amsmath,amsthm,amssymb,showpacs,floatfix,superscriptaddress,longbibliography]{revtex4-2}

\usepackage{dsfont}
\usepackage{xcolor}
\usepackage{float}
\usepackage{enumitem}
\usepackage{bbm}

\usepackage[normalem]{ulem}
\usepackage{siunitx}
\usepackage{graphicx}
\graphicspath{ {./images/} }

\usepackage[caption=false]{subfig} 

\usepackage{adjustbox}
\usepackage{dcolumn}
\usepackage{bm}
\usepackage[unicode=true]{hyperref}
\hypersetup{colorlinks=true,linkcolor=blue,filecolor=blue,urlcolor=blue,citecolor=blue}
\usepackage{nicefrac}
\usepackage{physics}
\usepackage{tikz}
\usetikzlibrary{arrows}
\usepackage{listings}

\newcommand{\CCC}[0]{$\mathds{C}^3$~}

\begin{document}

\title{
Benchmarking Optimization Algorithms for Automated Calibration of Quantum Devices
}

%=======================================================================================================

\author{Kevin Pack}
\affiliation{Peter Gr\"unberg Institut -- Quantum Computing Analytics (PGI 12), Forschungszentrum J\"ulich, D-52425 J\"ulich, Germany}

% \author{Miralem Sinanovic}
% \affiliation{Peter Gr\"unberg Institut -- Quantum Computing Analytics (PGI 12), Forschungszentrum J\"ulich, D-52425 J\"ulich, Germany}

% \author{Moritz Wald}
% \affiliation{Theoretical Physics, Saarland University, 66123 Saarbr\"ucken, Germany}

% \author{Anurag Saha Roy}
% \affiliation{Theoretical Physics, Saarland University, 66123 Saarbr\"ucken, Germany}
% \affiliation{Qruise GmbH, 66113 Saarbr\"ucken, Germany}

% \author{Nicolas Wittler}
% \affiliation{Peter Gr\"unberg Institut -- Quantum Computing Analytics (PGI 12), Forschungszentrum J\"ulich, D-52425 J\"ulich, Germany}

% \author{Federico Roy}
% \affiliation{Peter Gr\"unberg Institut -- Quantum Computing Analytics (PGI 12), Forschungszentrum J\"ulich, D-52425 J\"ulich, Germany}

\author{Shai Machnes}
\affiliation{Qruise GmbH, 66113 Saarbr\"ucken, Germany}
%\affiliation{Peter Gr\"unberg Institut -- Quantum Computing Analytics (PGI 12), Forschungszentrum J\"ulich, D-52425 J\"ulich, Germany}

\author{Frank K. Wilhelm}
\affiliation{Peter Gr\"unberg Institut -- Quantum Computing Analytics (PGI 12), Forschungszentrum J\"ulich, D-52425 J\"ulich, Germany}

%=======================================================================================================
%
\begin{abstract}
We present the results of a comprehensive study of optimization algorithms for the 
calibration of quantum devices. As part of our ongoing efforts to automate bring-up, 
tune-up, and system identification procedures, we investigate a broad range of 
optimizers within a simulated environment designed to closely mimic the challenges 
of real-world experimental conditions. Our benchmark includes widely used algorithms 
such as Nelder-Mead and the state-of-the-art Covariance Matrix Adaptation 
Evolution Strategy (CMA-ES). We evaluate performance in both low-dimensional 
settings, representing simple pulse shapes used in current optimal control protocols 
with a limited number of parameters, and high-dimensional regimes, which reflect the 
demands of complex control pulses with many parameters. Based on our findings, we 
recommend the CMA-ES algorithm and provide empirical evidence for its superior 
performance across all tested scenarios.
\end{abstract}

\maketitle

% \section{TODO}
% \begin{itemize}
%     % \item correct affiliations for everyone
%     \item Placeholder
% \end{itemize}

%=======================================================================================================
\section{Introduction}\label{sec:intro}
The advent of the Noisy Intermediate-Scale Quantum (NISQ) era has made operating quantum processing 
units (QPUs) increasingly complex, primarily due to the growing number of qubits~\cite{arute2019quantum, jurcevic2020demonstration}. One of the most 
time consuming and crucial parts is hereby the calibration of said device. As the performance of the 
tune up procedure is directly responsible for the final performance of the 
device and fidelity of the executed operations.
Calibration typically refers to the estimation and adjustment of essential device parameters, 
such as the intrinsic qubit frequencies or control-pulse characteristics (e.g., pulse amplitudes, 
durations, and drive frequencies), which must be carefully optimized to ensure reliable operation.

A key ingredient is the choice of the underlying classical optimization algorithm.
Therefore, we investigate a broad range of state-of-the-art optimization algorithms—particularly 
those developed in the field of machine learning—to identify a portfolio best suited for the 
calibration of quantum devices. 

This work presents the results of our comprehensive benchmarking effort in the context of 
system bring-up and calibration procedures, and is structured as follows:

We begin in Section~\ref{sec:problems} by examining the current challenges 
associated with the calibration of quantum devices. 
Section~\ref{sec:auto-calibration} introduces automated calibration as a potential 
solution to these challenges and discusses the difficulties that arise in its 
implementation. These difficulties motivate a set of criteria against which 
various optimization algorithms are evaluated.

Section~\ref{sec:algo-choice} provides an overview of optimization algorithms 
and outlines the rationale behind the selection of candidates considered suitable 
for our use case. Section~\ref{sec:benchmark} presents the results of our benchmarking 
experiments, highlighting characteristic behaviors, strengths, and weaknesses of each 
algorithm across different control pulse scenarios. 

Section~\ref{sec:discussion} offers a critical discussion of the benchmarking outcomes, 
and Section~\ref{sec:conclusions} concludes the work by summarizing the key findings. 
% Additional technical details, including implementation aspects of the algorithms, 
% are provided in the appendix. \Kevin{Appendix, yes, no?}
% \Kevin{Consider including an outlook on Moritz's paper here?}

\section{Challenges with Calibration of Quantum Devices}\label{sec:problems}
Currently, calibration of quantum processors is often done manually by experimentalists, typically 
involving the execution of a variety of experiments to determine 
the best parameters for hardware settings and control pulses to 
maximize operation fidelity.
Additionally, there is often a hierarchical structure in which 
these experiments are executed, as the precise determination of one 
parameter requires the prior knowledge of a different parameter. 
While one can argue that this process can be performed automatically by
a well-written suite of routines performing these experiments, the overall \textit{manual}
aproach to the calibration of the device does not change.
So far, this approach has given adequate results~\cite{Sheldon2016, Sheldon2016single}, yet there are some major 
problems associated with it.

First, manual calibration of quantum devices can take a considerable 
amount of time—often on the order of hours to weeks depending on the system 
and complexity. While this approach has historically yielded 
high-fidelity results, it can be susceptible to system drift. 
Temporal instabilities such as fluctuations in coherence times 
or shifts in control parameters may emerge during or between 
calibration stages, potentially degrading the effectiveness 
of the calibration if not mitigated~\cite{Proctor2020, Burnett2019}. 

Second, the approach is not applicable to each quantum device in the same way, 
as each device might need a different set of experiments to be calibrated. 
Furthermore, human time is spent on this task, which could be utilized in a 
more productive way as the calibration procedure does not deliver new scientific insights.
On the contrary, it has become a time-consuming hurdle to research.

Additionally, with the recent increase in the number of qubits this problem is amplified. 
The calibration of a device with fifty or more qubits can take weeks or months 
and require an immense number of personnel and therefore financial resources 
~\cite{arute2019quantum}. In summary, the scalability of this type of calibration is 
infeasible, especially looking ahead to quantum devices with thousands or millions of qubits.

Further, recent findings indicate that more complex control pulse shapes could lead to improved 
fidelities for gates~\cite{Werninghaus2021}. 
While complex control pulses may help to improve fidelities, they might result in an 
increase in control parameters to fine-tune during calibration.

\section{Automated Calibration}\label{sec:auto-calibration}
Therefore, we advocate an automated approach to the calibration to quantum processors to solve the aforementioned optimal control
problems~\cite{Kelly2014, Rol2017}. Here, an algorithm is used to optimize the operation we aim to implement on the device. 
To achieve this, the algorithm is provided with a \textit{goal function}—commonly referred 
to as a \textit{loss function} in the field of machine learning—which serves as a figure of merit. 
Ideally, this function incorporates all relevant characteristics of the target operation and 
provides a quantitative measure of its fidelity.
The notion of \textit{optimize} can refer to finding the minimum or 
maximum of a loss function. In the context of this paper we minimize the 
loss function.

An automated approach to calibration is not new and optimization 
algorithms -- most famously the Nelder-Mead algorithm -- have indeed been 
used in this context~\cite{Kelly2014, kelly2018dag, Rol2017}. In the past, the 
use of optimization algorithms was usually limited to the fine-tuning of 
the already calibrated control pulse. In our work we aim to utilize optimization 
algorithms to automate the process of calibration as a whole, contrary to the 
previous limitation to fine-tuning.
We are confident that this will reduce the time it takes to calibrate new devices
and improve the fidelities of device operations.

Ultimately, our approach to the calibration of a new device is encapsulated in the 
methodology of our software toolset \CCC~\cite{c3, c3_technical}, which extends beyond the calibration procedure. 
In particular, the methodology can be summarized as follows: The process begins by 
executing a minimal set of predefined experiments to estimate the minimal set of system parameters 
needed to create an initial model of the experimental system. We then use the model as a 
starting point for an open-loop optimization of a control pulse~\cite{Glaser2015, Machnes2011}. In the next step, the result of 
the open-loop optimization becomes the starting point of a closed-loop optimization.
During this stage, the experiment provides the optimization algorithm with feedback 
about the current fidelities of operations. This approach ensures that we are not limited 
by an imperfect model and allows us to fine-tune the control pulse in a model-free process. 
This procedure is known as \textit{Ad-HOC} and was originally proposed
in Ref.~\cite{Egger2014}. We extend the methodology by utilizing the experimental data of 
the closed-loop optimization to refine our model. This third and final stage 
is the model learning step. Here, we test our model in simulation against the 
collected experimental data during the closed-loop stage and potentially 
enhance the initial model.

For a more complete picture, we encourage the reader to consult our previous 
publications~\cite{c3, c3_technical} in which we demonstrate the full methodology 
in detail. Looking ahead, we note that \CCC will be succeeded by its forthcoming update, 
\textit{Paraqeet}.

\subsection{Challenges of Automated Calibration}\label{sec:challenges}
While automating the calibration process using optimization algorithms may 
appear straightforward, several challenges must be addressed to achieve 
successful device calibration.

One of the most important decisions is the choice of loss function, 
as the loss function ultimately governs the calibration 
process. The loss function fully incorporates 
characteristics of the desired operations. As such, the loss function should 
serve as a figure of merit for the desired operations, or as an aggregation 
of multiple condensed into a single scalar value.
In the past the Randomized Benchmarking protocol~
\cite{Magesan2011, Knill09, Chow2009, Magesan2012, Corcoles2013}, 
Quantum Process Tomography~\cite{QPTCNOT, Chow2009} and its variants, or 
ORBIT~\cite{ORBIT} have been used in this process. 
In particular, ORBIT has been designed for the task of automated calibration.

A priori, there is no \textit{best} loss function, any figure of merit that 
measures the fidelity of the executed operation is fit to be a loss function.
As such we are free to choose any figure of merit available to us, which 
encapsulating the properties we are interested in most or is most suitable
for the current situation.

The choice of the loss function can directly impact the likelihood of successful calibration. 
The optimization landscapes associated with different loss functions may have
beneficial or detrimental properties for the chosen optimization algorithm~\cite{zhu2022benchmarking}.

Throughout this work, we collected several performance-defining 
aspects of automated calibration and refined them into a list of criteria.
The optimization algorithms are benchmarked against these criteria to 
gauge the suitability of the different algorithms for the calibration task.

\begin{figure*}[t]
    \centering
    \includegraphics[width=1.0\linewidth]{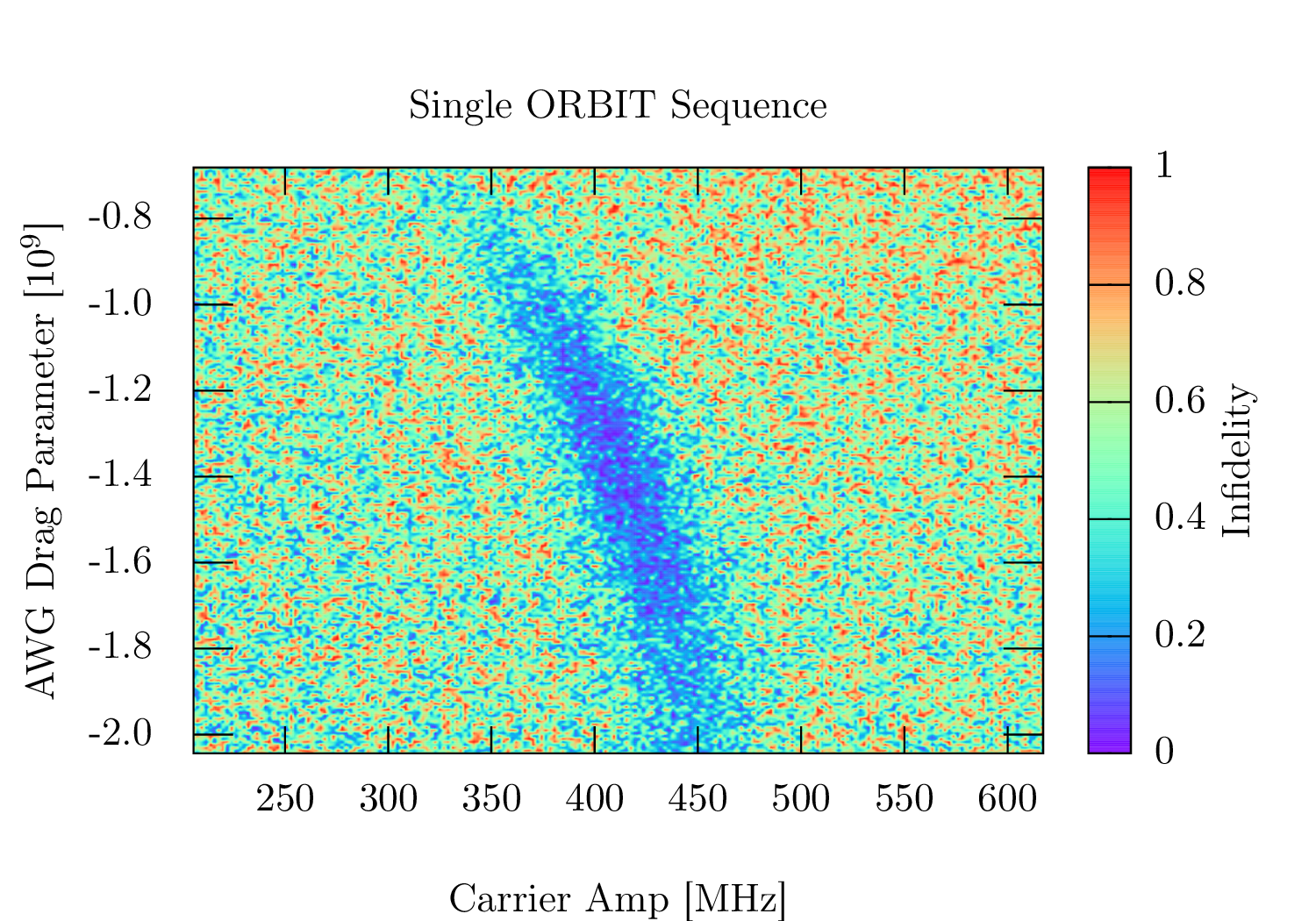}
    \caption{Simulated landscape of loss function specified in Eq.~(\ref{eq:loss_infid_orbit}) for a single ORBIT sequence $n=1$ and $l = 80$ 
    for a DRAG pulse. The typical DRAG minimum is clearly visible. The landscape exhibits 
    multiple local minima, which can hinder the calibration process, as 
    optimizers may converge to sub optimal regions. Averaging over a 
    sufficiently large number of ORBIT sequences helps smooth out these 
    features and improves the overall convergence behavior of the optimizer.}
    \label{fig:local-minima}
\end{figure*}

In the following, we present the aforementioned criteria:
\begin{itemize}
    \item   \textbf{Noise Resistance}
    
            Real-life experimental data is by nature noisy, as such, the 
            loss function will inherit this property. Additionally, the 
            quantum nature of the experiment will on top of that also 
            exhibit sampling noise.
            An optimization algorithm needs to be able to converge reliably
            despite noise.
    \item   \textbf{Ability to Escape Local Extrema}
    
            The choice of the loss function impacts the optimization landscape. 
            These landscapes contain multiple local extrema, as seen in
            Fig~\ref{fig:local-minima}. As high fidelity operations are required,
            the convergence to the global optimum is desired. 
            Overcoming local extrema is therefore imperative during 
            the convergence process.
    \item   \textbf{Dimension Scaling}
    
            The number of parameters needed to operate quantum devices may rise
            with an increase in complexity in the hardware or with 
            more complex control pulses. As such, any potential algorithm is required
            to successfully converge even in a high-dimensional setting.
    \item   \textbf{Convergence Budget/Speed of Convergence}
    
            As experimental time is valuable, the calibration process should be kept 
            as short as possible to maximize device. Furthermore, calibration must 
            be completed on a timescale shorter than that of the system's drift.
    \item   \textbf{Batching}
    
            Many experimental setups allow the upload of a batch of waveforms 
            to the connected devices to speed up the execution time of
            an experiment. As such, the optimization algorithm is expected 
            to test different settings in a single iteration by supplying
            multiple potential candidate solutions.
    \item   \textbf{Ease of Setup: Hyperparameters}
    
            Each optimization algorithm approaches the task of optimizing the 
            loss function using different methodologies and hence has its own 
            set of internal algorithmic parameters. These internal parameters are 
            commonly called the \textit{hyperparameters}. For 
            successful convergence of the algorithm, an initial hyperparameter set 
            is often required. The ease of determining these parameters and the 
            number of hyperparameters can therefore be crucial for a specific 
            optimization task. This can be complicated by the fact that, 
            due to the nature of the task, we have limited 
            knowledge of the system being optimized.
\end{itemize}
Hyperparameter settings, in particular, have been shown to be essential 
for successful convergence of the algorithm~\cite{Probst2019, Weerts2020}. It remains a challenge to 
transform limited prior knowledge of the system or the experience of a 
skilled experimentalist into suitable starting values for these parameters.

\section{Choice of (Gradient-Free) Algorithms}\label{sec:algo-choice}
In general, there are two types of optimization algorithms: gradient-free 
algorithms and gradient-based algorithms. As indicated by the classification,
the distinction is based on whether the algorithm utilizes the gradient 
of the loss function.

By nature, a calibration process does not allow us to analytically 
access the gradient of the loss function, as we do not have a detailed model.
As such, we are limited to gradient-free optimization algorithms. These types of algorithms 
usually take more iterations to converge, as the availability of a gradient allows for much 
faster convergence. This limitation is offset by the fact that gradient-free algorithms 
can be applied in situations where either an incomplete model or no model is available. 
An optimization task with limited or no knowledge of the system is usually referred to as
\textit{black-box optimization}.

Research on gradient-free optimization algorithms is a vast and active field, 
with hundreds of black-box optimization methods in the literature.
These include Evolutionary Strategies (ES)~\cite{LI2020100694}, Particle Swarm Optimization (PSO)~\cite{kennedy95pso, bratton2007defining}, 
Differential Evolution (DE)~\cite{Kochenderfer2019}, Random Search, Simultaneous Perturbation 
Stochastic Approximation (SPSA)~\cite{WangChen2020}, the Nelder–Mead method~\cite{Nelder-Mead, Nelder-Mead-slow-dim}, Bayesian Optimization~\cite{7352306}, and others. 

Following an initial review—particularly focused on algorithms used in the field 
of machine learning, where reliable optimization is of high priority—we narrowed 
the list to the following candidates:
\begin{itemize}
    \item   \textbf{Covariance Matrix Adaptation Evolution Strategy (CMA-ES)}

            The covariance matrix adaptation evolution strategy, or, for short, 
            CMA-ES~\cite{Hansen2003, hansen2023cmaevolutionstrategytutorial}, 
            is part of the class of \((\mu/\rho,\lambda)\) or \((\mu/\rho + \lambda)\) 
            evolution strategy algorithms. 
            The algorithm is regarded as state-of-the-art in the field of 
            gradient-free optimization.
            
            \textit{Hyperparameters: population size, initial standard deviation 
            \(\sigma\) of the distribution}

    \item   \textbf{Powell's Method}

            Powell's method~\cite{Powell1964, Scipy} is classified as a type 
            of intelligent search algorithm. It systematically explores each 
            parameter direction and updates the search direction based on the 
            current search state.
            
            \textit{Hyperparameters: None}

    \item   \textbf{Nelder–Mead}

            The Nelder–Mead algorithm~\cite{Nelder-Mead, Nelder-Mead-slow-dim} 
            is one of the oldest optimization algorithms but is still commonly 
            used for low-dimensional or unconstrained optimization tasks. 
            The algorithm has been applied to calibration problems in the past 
            and serves as a baseline in this study.
            
            \textit{Hyperparameters: initial standard deviation \(\sigma\) of 
            the distribution}

    \item   \textbf{1+1-ES}

            The \(1+1\)-ES~\cite{nevergrad} is another \((\mu/\rho + \lambda)\) 
            evolution strategy algorithm. While the algorithm is relatively simple, 
            it has proven effective in initial evaluations.
            
            \textit{Hyperparameters: initial standard deviation \(\sigma\) of the 
            distribution}

    \item   \textbf{Differential Evolution (DE)}

            The Differential Evolution algorithm~\cite{Kochenderfer2019}, like 
            \(1+1\)-ES and CMA-ES, belongs to the class of evolution strategies. 
            A population of candidate solutions is modified in each iteration, 
            mimicking evolutionary processes by perturbing the individual 
            vector elements of each candidate.
            
            \textit{Hyperparameters: initial standard deviation \(\sigma\) of 
            the distribution, population size, crossover parameter, 
            differential weight parameter}

    \item   \textbf{Simulated Annealing (SA)}

            Simulated Annealing~\cite{Kochenderfer2019} is inspired by the 
            ability of quantum-mechanical systems to escape local minima through 
            tunneling. In the algorithm, this is achieved by accepting worse 
            solutions with nonzero probability to potentially find better solutions.
            
            \textit{Hyperparameters: temperature decay schedule, initial temperature, 
            temperature decay rate \(\gamma\), initial standard deviation \(\sigma\) 
            of the distribution}
\end{itemize}

% \Kevin{do we want some explanatory text to each algorithm? I am not sure 
% if this doesn't blow up too much but it also adds some context}

The above-mentioned algorithms have first been tested on different 
analytical loss functions, each representing distinct challenges 
that may arise during the calibration of a real-life experiment. 
These types of benchmarks, evaluating optimization algorithms on analytical loss functions, 
are widely available and have been conducted by various research efforts. 
As such, we include these benchmarks in the appendix for completeness, 
while shifting the focus of this work to benchmarks performed in simulation.

\section{Benchmarking in Simulation}\label{sec:benchmark}
As we are interested in the performance of the chosen algorithms for
the task of calibration, we decided to simulate a single-qubit experiment.
In simulation, we calibrate a single qubit and utilize ORBIT as the loss function.
The ORBIT protocol can be summarized as follows: Create a sequence \(s\)
of length \(l\) composed of random Clifford gates \(C_i\) for \(i = 1,\dots, l-1\), 
which are elements of the Clifford group, and finally reverse the sequence 
with a single Clifford gate \(C^{-1}\). The properties of the Clifford group 
guarantee the existence of such a gate \(C^{-1}\).
Finally, we measure the final state after the execution of the ORBIT sequence. 
If our operations were without error, the measured state would be the ground state. 
As our real-life gates are not perfect, the measurement of the final system state provides 
a quantitative figure of merit for the fidelity of our operations. 
Ideally, this process is repeated for \(N\) independently sampled ORBIT sequences,
as averaging over multiple sequences provides a more stable figure of merit.
Since we are interested in minimizing gate errors, 
we calculate the infidelity of our performed operations. 
Finally, we take the logarithm of the infidelity to enhance sensitivity 
to small changes in the low-infidelity regime.

In summary, our loss function \(l(\bm{x})\) can be mathematically described as:
\begin{equation}
    l(\bm{x}) = \ln\left(1 - \frac{1}{N} \sum_n \mathcal{M}(\hat{s}_n(\bm{x}))\right)
    \label{eq:loss_infid_orbit}
\end{equation}
with the measurement \(\mathcal{M}(\hat{s}_n(\bm{x}))\) defined as
\begin{equation}
    \mathcal{M}(\hat{s}_n(\bm{x})) = |\bra{0}\hat{s}_n(\bm{x})\ket{0}|^2
\end{equation}
and the ORBIT sequence \(\hat{s}_n\) of length \(l\) given by
\begin{equation}
    \hat{s}_n(\bm{x}) = \left(\prod_i^{l-1} \hat{C}_i(\bm{x})\right)\hat{C}^{-1}(\bm{x}).
\end{equation}
Here, \(\bm{x}\) represents the current set of pulse parameter settings provided by the 
optimization algorithm, and the executed gates are therefore dependent on \(\bm{x}\).

Alternatively, we can also choose \(C^{-1}\) such that the ORBIT sequence ends in the first 
excited state rather than the ground state. This can help mitigate local minima in the 
optimization landscape, where a small pulse amplitude may cause the system to remain in 
the ground state. In this case, the measurement must target the excited state \(\ket{1}\) 
instead of the ground state \(\ket{0}\).

For each evaluation of the loss function, a new set of random ORBIT sequences is generated.
This prevents the optimizer from overfitting to a fixed set of circuits and ensures that
the loss function approximates the expectation value over random Clifford sequences.

The stochastic nature of the loss function is determined by the number of ORBIT
sequences \(N\) averaged over. In an experimental setting, \(N\) is often limited
by the execution time required to evaluate the loss function. Choosing \(N\)
too large can drastically increase the time required to calibrate the device.
As such, automated calibration with ORBIT as a loss function is inherently
a stochastic optimization problem. However, since the optimizer evaluates
many different randomly sampled sequence sets during the optimization,
the resulting parameters are not tied to a particular sampled sequence set.

Additionally, given that ORBIT is intended to speed up the calibration 
of quantum devices by utilizing techniques from RB without requiring the full, time-intensive 
nature of the protocol, it is worth examining the implications that the RB figure of 
merit, and thus, by extension, ORBIT, has for the characterization of the performance of the calibrated gates. 
It was shown in \cite{Wallman_2014} that one can derive theoretical bounds relating the RB error rate to 
worst-case error measures. Furthermore, \cite{Hashim2023} showed that while these worst-case estimations are often 
smaller in experiment than predicted by theory, they can still differ significantly from the 
average error rates reported by RB. 
Therefore, it should be noted that ORBIT, as a figure of merit during calibration, does not guarantee uniform 
gate set performance, but should rather be interpreted as indicating good average 
performance with respect to the sampled ORBIT sequences.

Finally, while ORBIT is closely related to randomized benchmarking,
it does not offer the same SPAM robustness in the evaluation of the
figure of merit. This is because ORBIT does not perform the exponential
decay fitting used in randomized benchmarking to separate gate errors
from SPAM contributions. Instead, SPAM errors contribute approximately
as a constant offset to the measured survival probability, since
state-preparation and measurement errors affect all sequences in a
similar manner and are largely independent of the pulse parameters
being optimized.

\subsection{The Simulated System Setup}
For the simulated system, we chose a three-level system modeled by a Duffing oscillator 
driven by an X-quadrature. The full system Hamiltonian is given by
\begin{equation}
    H = \omega\hat{a}^\dag \hat{a}  +  \frac{\delta}{2} (\hat{a}^\dag \hat{a} - 
    \hat{\mathds{1}})\hat{a}^\dag \hat{a} + c(t)(\hat{a}^\dag + \hat{a}),
\end{equation}
where \(\omega\) is the qubit frequency, \(\delta\) is the anharmonicity, \(c(t)\) is the 
control field, and \(\hat{a}\)/\(\hat{a}^\dag\) are the lowering and raising operators, respectively.

The qubit frequency was set to 4.8 GHz, and the anharmonicity to 200 MHz.

To benchmark the list of optimizers against different requirements, we use two 
types of pulses: a derivative removal by adiabatic gate (DRAG) pulse~\cite{Motzoi2009} 
and a piecewise constant (PWC) pulse. The DRAG pulse is a standard choice 
in experimental setups due to its nature of reducing leakage into higher states~
\cite{Motzoi2009, Chen2016}. While it provides a good reference, it is a relatively simple pulse 
with a small number of tunable parameters.

As shown in~\cite{Werninghaus2021}, the simplicity of these pulses also limits the achievable 
fidelities. Therefore, we include benchmarks using a piecewise constant pulse, 
which typically involves many tunable parameters.

Hence, the two pulse types allow us to evaluate the optimizers 
in both a low-dimensional setting (DRAG) and a high-dimensional setting (PWC).

\subsubsection{Derivative Removal by Adiabatic Gate (DRAG) Pulse}
The input signal of the DRAG pulse is given by the following equation,
derived from the derivative removal by adiabatic gate (DRAG) method:
\begin{equation}\label{eq:Gaussian-with-DRAG}
  \begin{aligned}
    \varepsilon(t)  &= A\,\Omega_\text{Gauss}(t)\,\cos(\omega_d t + \phi_{xy})\\
            &+ \frac{1}{\delta}\,A\,\dot{\Omega}_\text{Gauss}(t)\,
                \sin(\omega_d t + \phi_{xy}).
  \end{aligned}
\end{equation}
Here, \(A\) is the amplitude of the pulse, \(\Omega_\text{Gauss}(t)\) is a Gaussian envelope, 
\(\omega_d\) is the drive frequency, \(\delta\) is the so-called DRAG parameter, 
and \(\phi_{xy}\) is the phase of the pulse. The phase can be used to implement 
different single-qubit gates with the same pulse shape~\cite{Motzoi2009, Chen2016}.

\subsubsection{Piecewise Constant (PWC) Pulse}
For the Piecewise Constant Pulse (PWC), the previous envelope \(\Omega_\text{Gauss}\) 
is replaced by a piecewise constant envelope \(\Omega_\text{PWC}\). In this setup, to 
provide a working initial guess, the shape of the step function is chosen as a 
discretization of \(\Omega_\text{Gauss}\). The optimization task for the PWC pulse 
lies in the fine-tuning of each individual value for the step 
height of the envelope. In order to benchmark the optimization
algorithms in a high-dimensional setting, the number of steps is chosen to be 41. 
As the pulse is based on the previous DRAG solution, we not only need to 
discretize \(\Omega_\text{Gauss}\) but also the DRAG correction, which depends on 
\(\dot{\Omega}_\text{Gauss}\). As such, the number of parameters to optimize doubles, 
resulting in a total of 82 parameters with the inclusion of the step values 
for \(\dot{\Omega}_\text{PWC}\).

\begin{figure*}[t]
    \centering
    \includegraphics[width=1.0\linewidth]{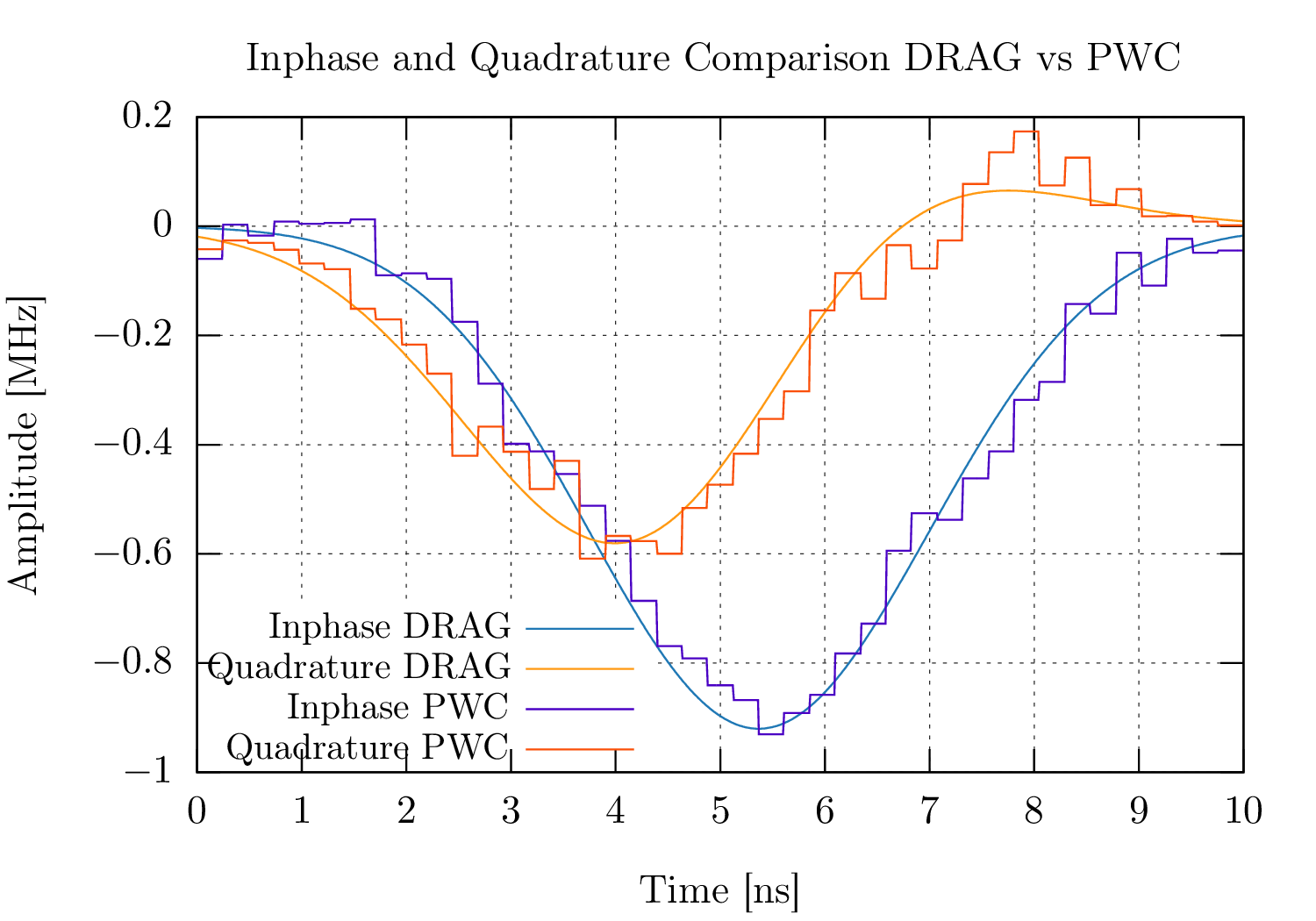}
    %\resizebox{0.8\textwidth}{!}{\input{images/inphase_quadrature_drag_pwc.tex}}  % Adjusted path for the .tex file     
    \caption{In-phase and quadrature components of the DRAG and PWC pulses for an 
    \( R_x(\theta = \frac{\pi}{2}) \) gate. The PWC pulse exhibits the 
    characteristic discretized step structure, where each step value is 
    independently optimized and forms part of the 82-dimensional optimization 
    space. At the start of the optimization, the step heights are randomized to 
    simulate an uninformed initial guess.}
    \label{fig:drag-pwc-pulse-comp}
\end{figure*}

The difference between a DRAG and PWC pulse is shown in Fig.~\ref{fig:drag-pwc-pulse-comp}.
The figure shows the in-phase and quadrature components of the simulated signal. 
One can see the PWC step function as a discretization of the DRAG signal. The mismatch 
of step height to the original DRAG signal stems from an introduced detuning of $5~\%$ 
to each step value. This detuning guarantees that the starting position of the PWC pulse
is not too close to the minimum found by the DRAG pulse.

\subsection{Hyperparameter Optimization}
Hyperparameters are the individual parameters of the optimization algorithm itself. 
These parameters can be very different in nature and sometimes rather abstract; 
in particular, the latter often makes it difficult to develop an 
intuition for optimal settings.
While there are algorithms that update their hyperparameters at each iteration, 
other algorithms do not and have static hyperparameters. If the hyperparameters 
of an algorithm are detuned, convergence can be hindered, and in the worst case, 
the algorithm may fail to converge entirely.

To ensure the validity of our benchmarks, we began by optimizing the hyperparameters 
of each individual algorithm. This was accomplished by running a series of full 
optimization procedures in simulation, where the hyperparameter settings for each 
run were selected by the CMA-ES algorithm. 

The loss function used for this hyperparameter optimization rates the 
performance of the optimizer with respect to the chosen hyperparameter 
configuration. It consists of two weighted terms:

First, we compute the slope of a first-degree polynomial fitted to 
the logarithmic values of all recorded loss function measurements 
during the optimization run with the current hyperparameter settings. 
This slope describes how quickly the algorithm converges — the greater
the negative value of the gradient, the faster the convergence. 
The second term of the rating is the logarithmic mean of the final $l$ loss 
function values obtained during the run. This captures information about 
the overall quality of the solution reached.
To obtain a more robust performance estimate, we repeat the optimization $n$ 
times with different random seeds for the same hyperparameter settings 
and average the individual ratings across runs.

Accordingly, we define our hyperparameter rating function \( r \) as follows:
\begin{equation}
    r = \sum_n r_n = \sum_n s\cdot m_n + \gamma \cdot e_n
    \label{eq:hyperparameter_rating_func}
\end{equation}
with \( e_n \) denoting the logarithm of the mean of the last \( l = 50 \) loss function values:
\begin{equation}
    e_n = \ln\left(\mu_n = \frac{1}{l} \sum_{i=k - l}^k f(x_{in})\right)
\end{equation}
and \( m_n \) representing the gradient of a first-degree polynomial fitted through 
the set of data points \( X_n \) and \( Y_n = \ln(f(X_n)) \). The parameters \( s \) 
and \( \gamma \) are weighting coefficients, set to \( s = 100 \) and \( \gamma = 1.0 \), 
respectively. The number of seeds used for averaging was set to \( n = 120 \).

\begin{figure*}[t]
    \centering

    % Subfigure 1
    \subfloat[Development of the overall score during hyperparameter optimization. 
    The gradual descent reflects improvements in convergence of the underlying 
    optimization process with each new hyperparameter setting.
    ]%
    {%
        \includegraphics[width=0.46\textwidth]{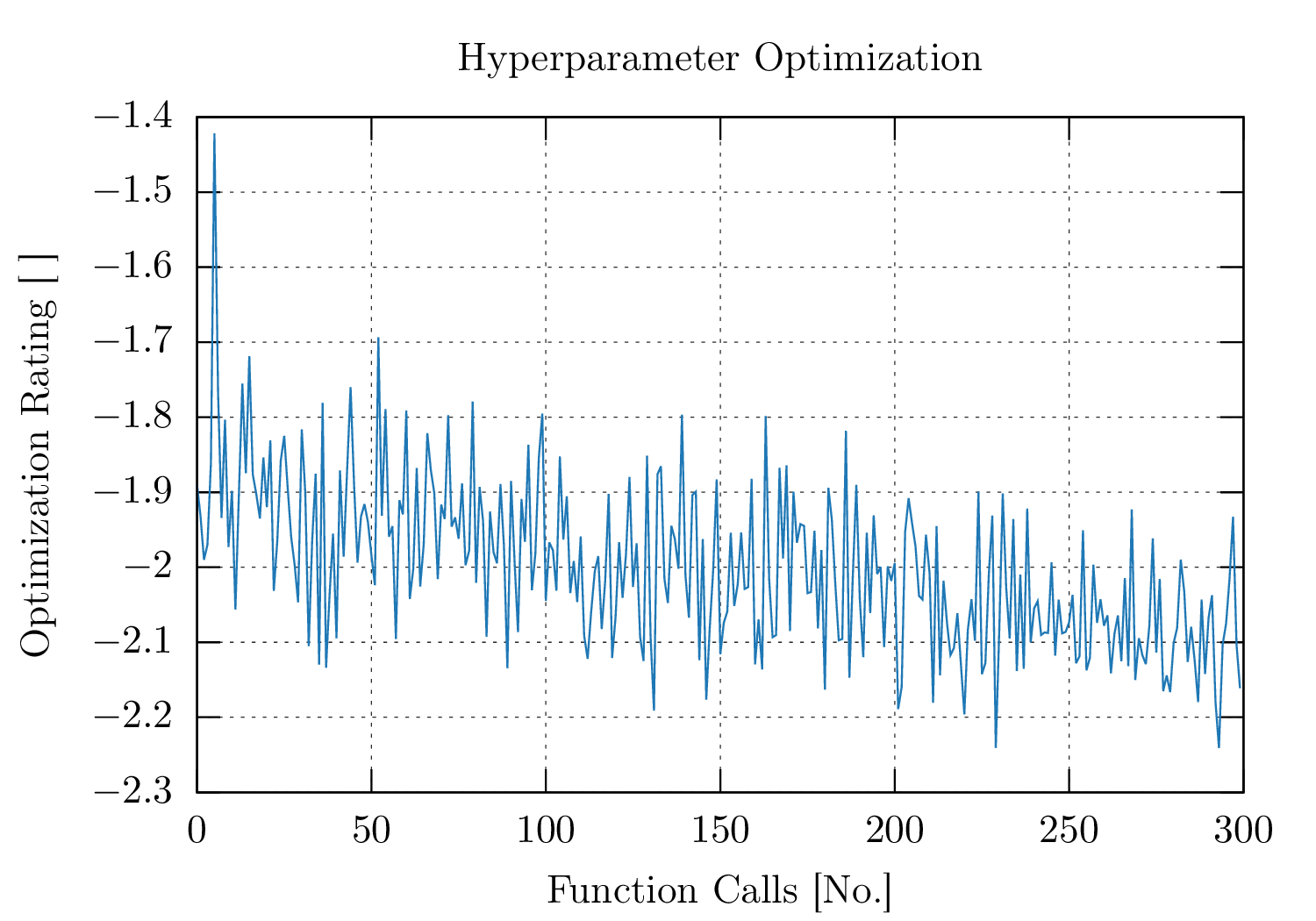}
        \label{fig:drag_cmaes_hyperparam_benchmark}
    }
    \quad
    % Subfigure 2
    \subfloat[Evaluated hyperparameter configurations by the CMA-ES algorithm. 
    Each point represents a tested combination of the initial standard deviation and population size. 
    The landscape shows how certain configurations lead to reduced optimization performance.
    ]%
    {%
        \includegraphics[width=0.47\textwidth]{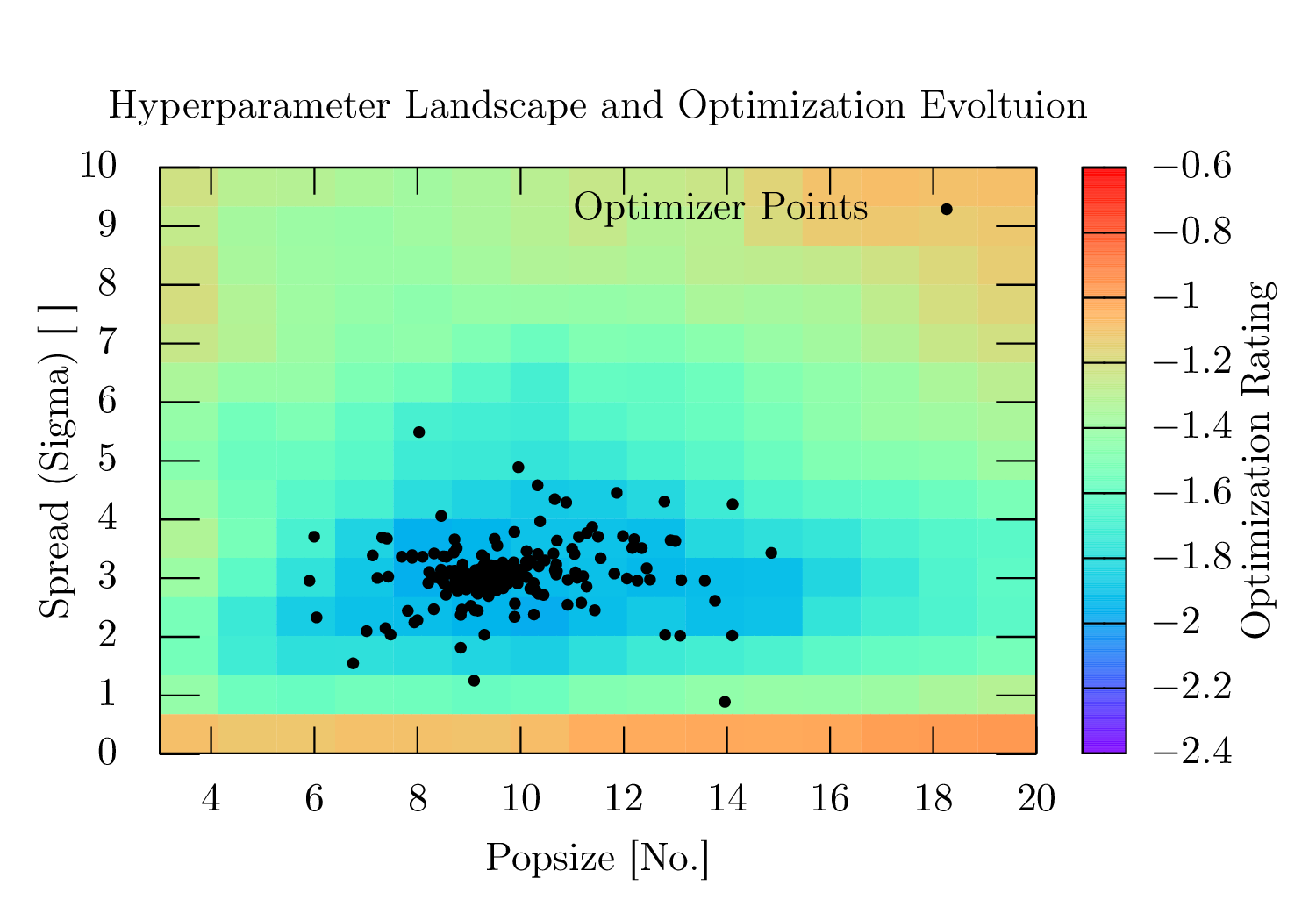}
        \label{fig:drag_cmaes_hyperparam_scatter}
    }

    % Main caption
    \caption{Hyperparameter optimization using the CMA-ES algorithm. 
    Panel~\protect\subref{fig:drag_cmaes_hyperparam_benchmark} shows the progression of the overall optimization score, 
    which combines final infidelity and convergence rate. 
    Panel~\protect\subref{fig:drag_cmaes_hyperparam_scatter} shows the history of evaluated hyperparameter values, 
    illustrating how some configurations degrade performance.}
    \label{fig:hyperparam_optim}
\end{figure*}

Figure~\ref{fig:hyperparam_optim} shows exemplary such a hyperparameter optimization for the 
CMA-ES algorithm. In Fig.~\ref{fig:drag_cmaes_hyperparam_benchmark} the overall convergence 
of the optimization can be seen through the decline of the loss function with growing 
function call number. Additionally, Fig.~\ref{fig:drag_cmaes_hyperparam_scatter} visualizes 
the evolution of the population of the optimizing CMA-ES instance on the loss landscape 
described by Eq.~\ref{eq:hyperparameter_rating_func}.

\begin{figure*}[t]    
    \centering
    \includegraphics[width=\textwidth]{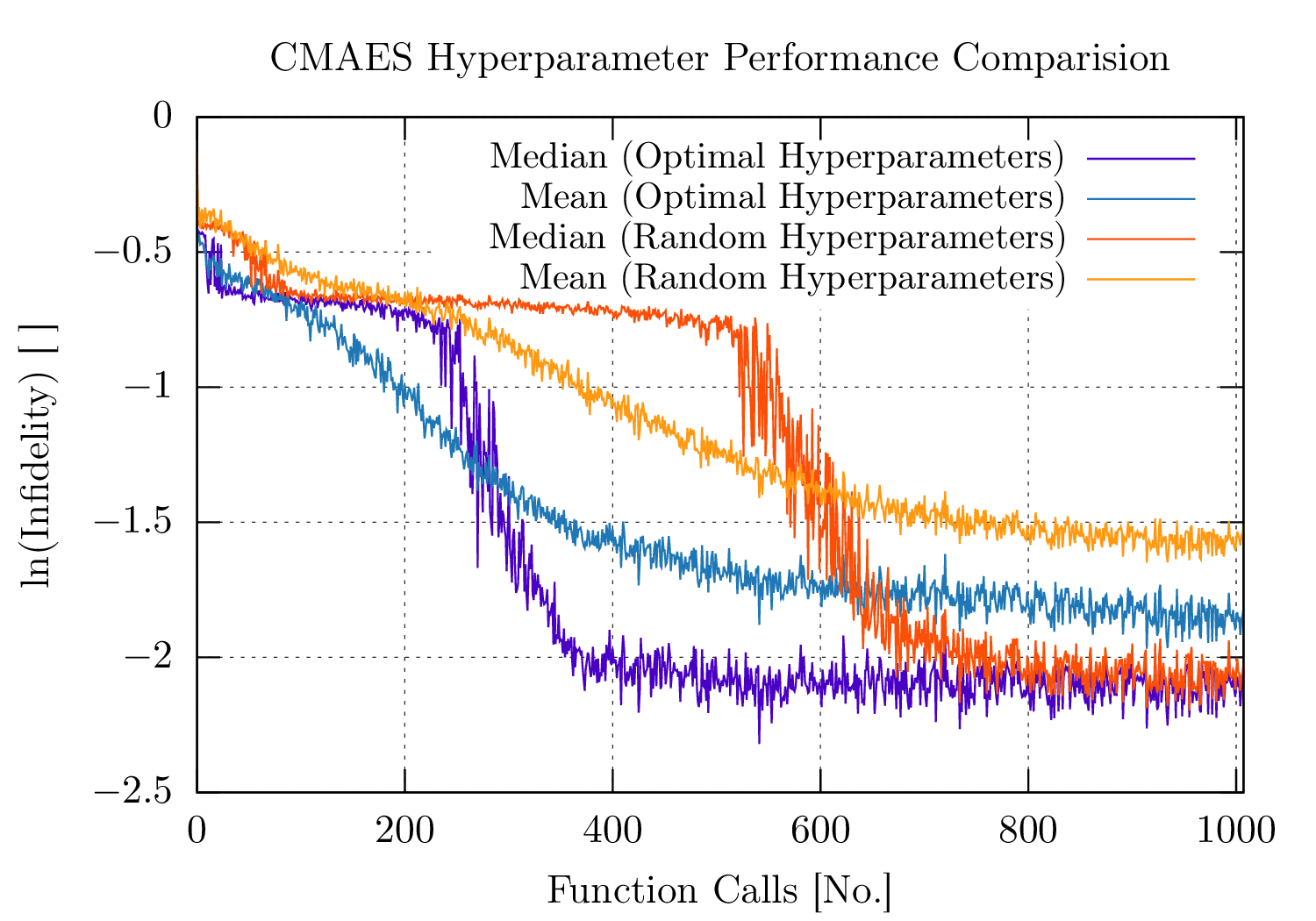}
    %\resizebox{0.8\textwidth}{!}{\input{images/drag_cmaes_hyperparam_comp_mean_median.tex}}  % Adjusted path for the .tex file    
    \caption{Comparison of 120 CMA-ES optimization runs using optimized hyperparameters versus randomly detuned hyperparameters. The figure shows the mean and median performance of the instances. The results demonstrate that CMA-ES performs better and converges with fewer function evaluations when hyperparameters are optimized.}
    \label{fig:drag_cmaes_hyperparam_comp_mean_meadian}
\end{figure*}

Finally, Fig~\ref{fig:drag_cmaes_hyperparam_comp_mean_meadian} shows a comparison
between CMA-ES optimizations with optimized hyperparameters compared to 
randomly chosen hyperparameters. While both converge with increasing numbers of 
function calls it can be seen that the CMA-ES instances with optimized hyperparameters
do converge in less function calls.

\subsubsection{Realistic Starting Position}
To make the simulation as realistic as possible, the initial detuning of the 
pulse parameters from their fine-tuned values must be defined. We chose a 
worst-case scenario in which each parameter is initially detuned by \(5\%\) 
from its optimal value.

This detuning is intended to mimic the state of the system after a rough 
calibration has been performed, representing the starting point for automated 
calibration using one of the tested algorithms to continue the QPU tune-up procedure.

\subsection{Benchmarking}

In the following comparisons, we present the performance of the algorithms 
introduced in section~\ref{sec:algo-choice}, evaluated in both a low-dimensional 
setting (DRAG pulse) and a high-dimensional setting (PWC pulse).

\begin{figure*}[t]
    \centering
    \subfloat[Mean performance of the different algorithms over 120 optimization 
    runs with varying random seeds. A decrease in infidelity indicates convergence. 
    Differential Evolution performs the worst, while CMA-ES achieves the 
    lowest overall infidelity.
    ]
    {%
        \includegraphics[width=0.475\textwidth]{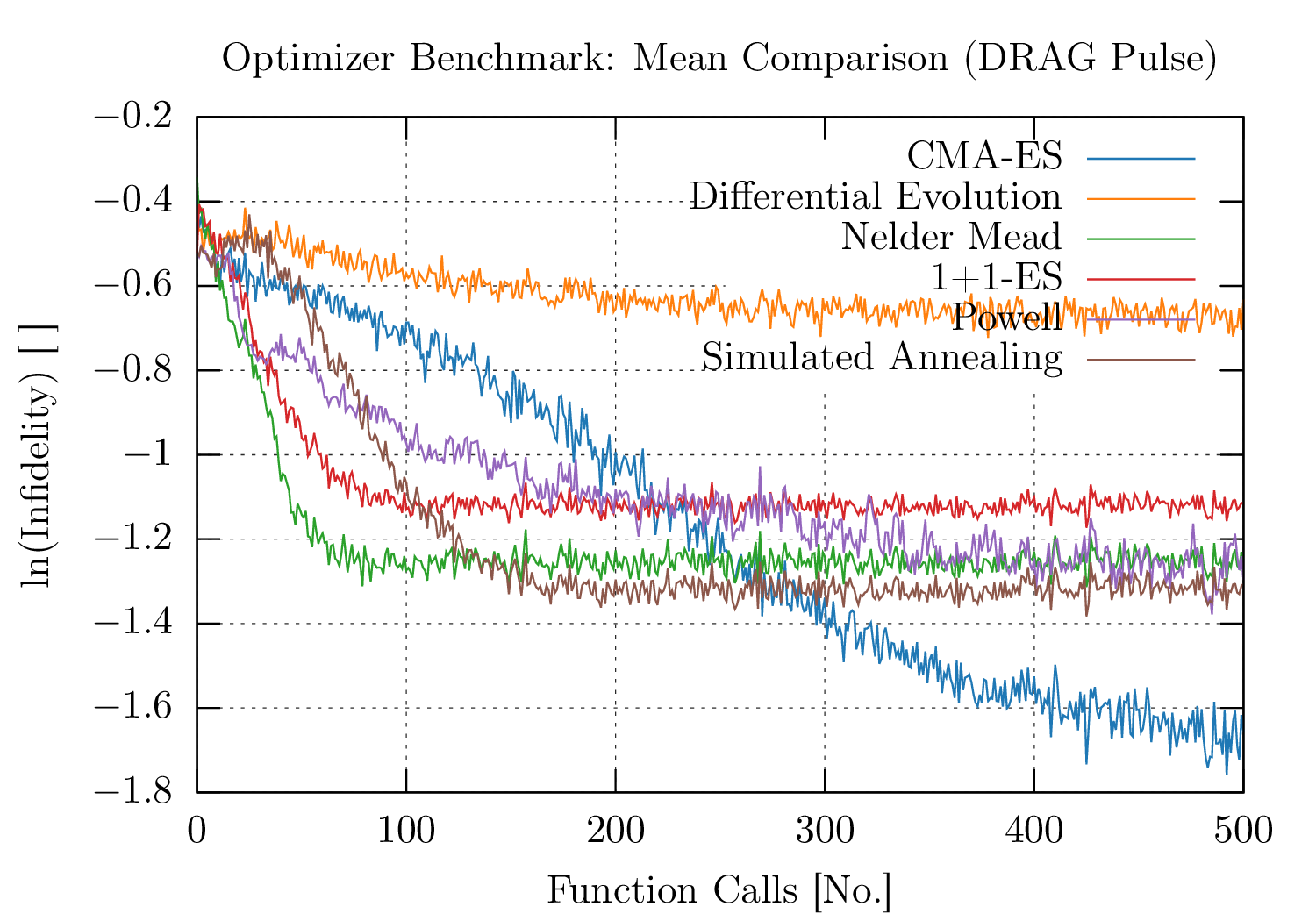}
        \label{fig:drag_mean_comp}
    }
    \quad
    \subfloat[Median performance of the different algorithms over 120 seed instances. 
    As with the mean, a decrease in infidelity indicates convergence. 
    CMA-ES consistently outperforms the other algorithms.
    \vspace{0.45cm}
    ]
    {%
        \includegraphics[width=0.475\textwidth]{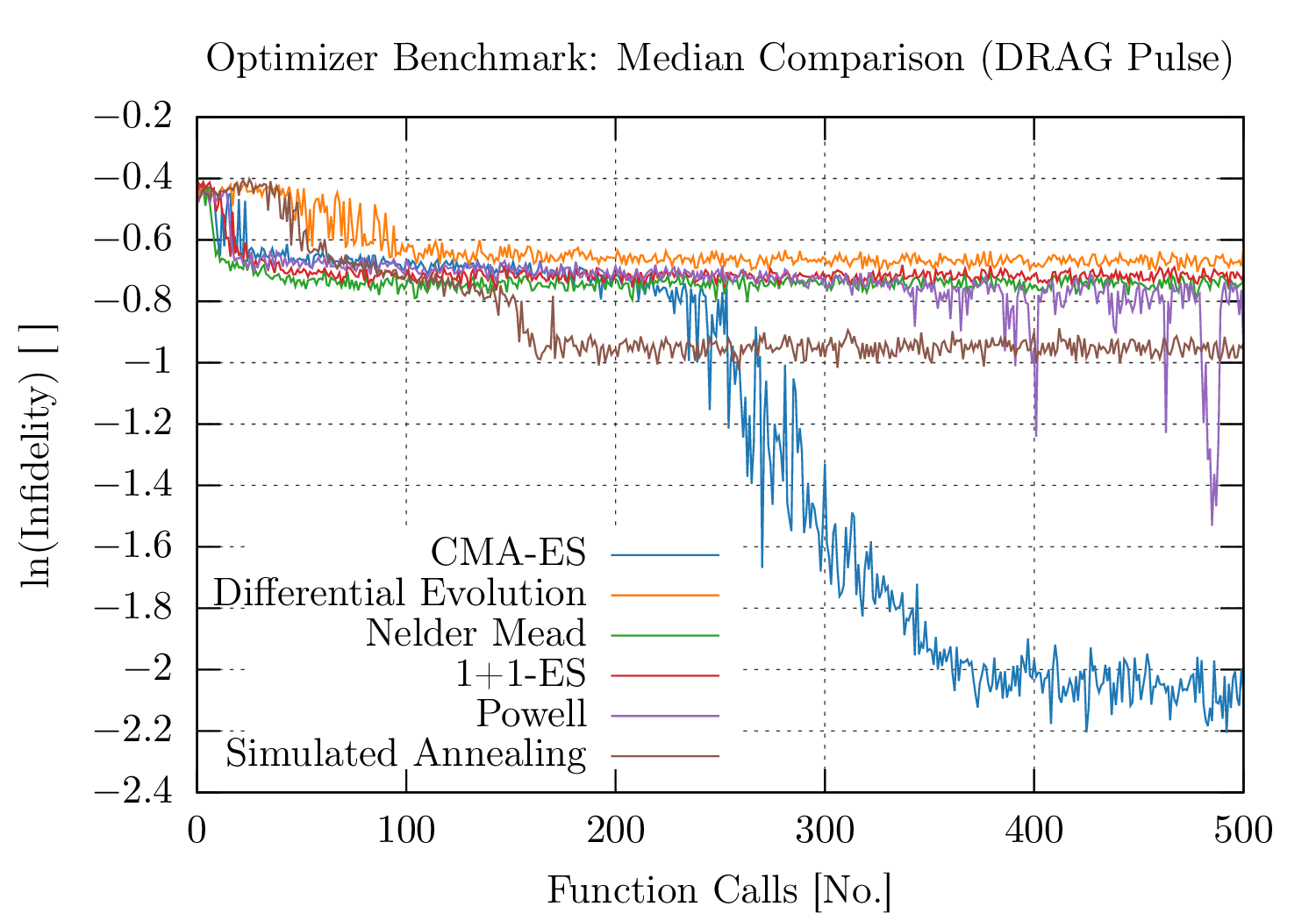}
        \label{fig:drag_median_comp}
    }

    \vspace{0.5cm}

    \subfloat[Mean performance of the algorithms for the PWC pulse over 120 seeded runs. 
    CMA-ES again achieves the lowest overall infidelity, while Differential Evolution 
    performs the worst.
    ]
    {%
        \includegraphics[width=0.475\textwidth]{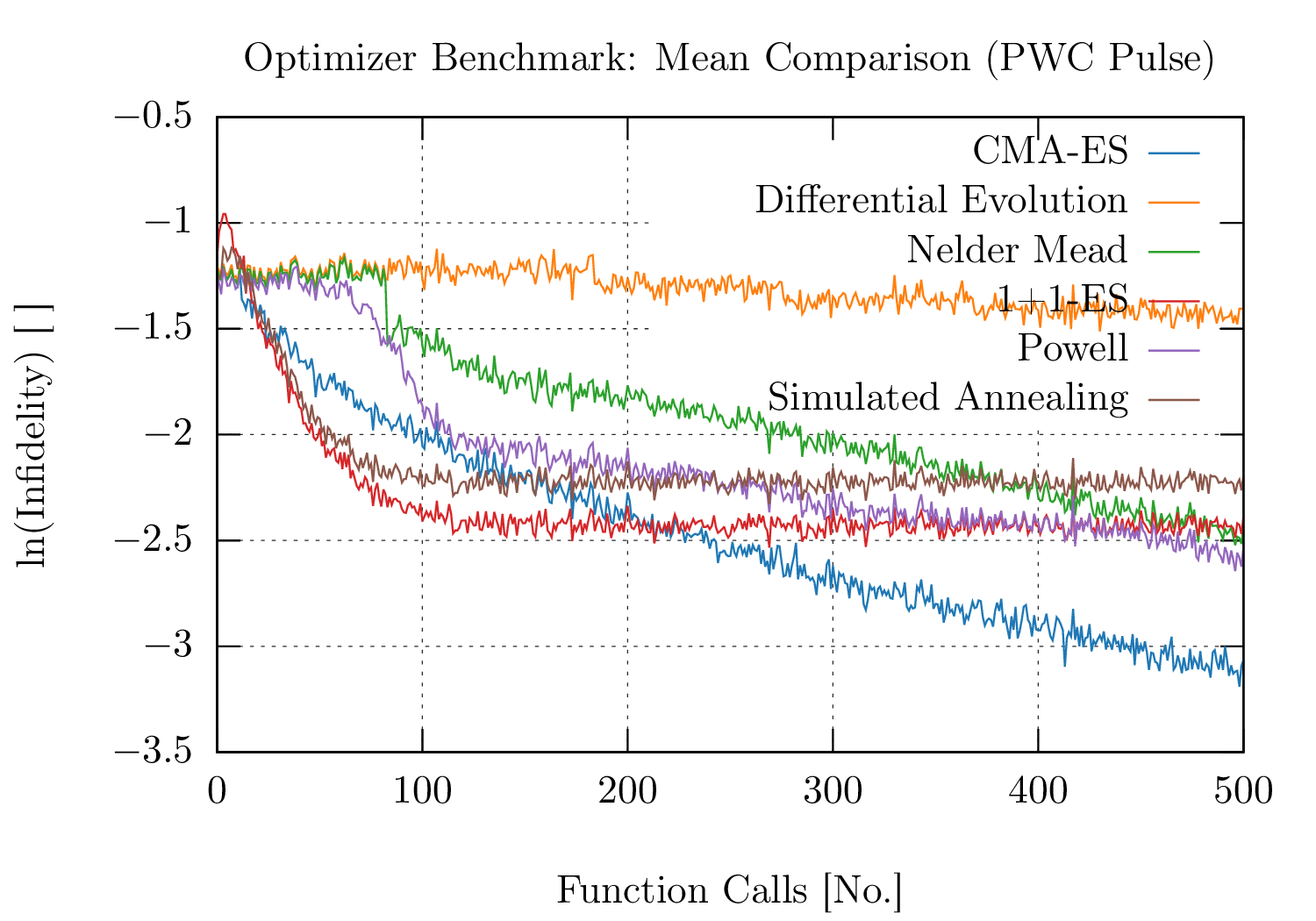}
        \label{fig:pwc_mean_comp}
    }
    \quad
    \subfloat[Median performance of the algorithms for the PWC pulse. 
    CMA-ES maintains the best performance across runs, indicating consistent convergence.
    \vspace{0.45cm}
    ]
    {%
        \includegraphics[width=0.475\textwidth]{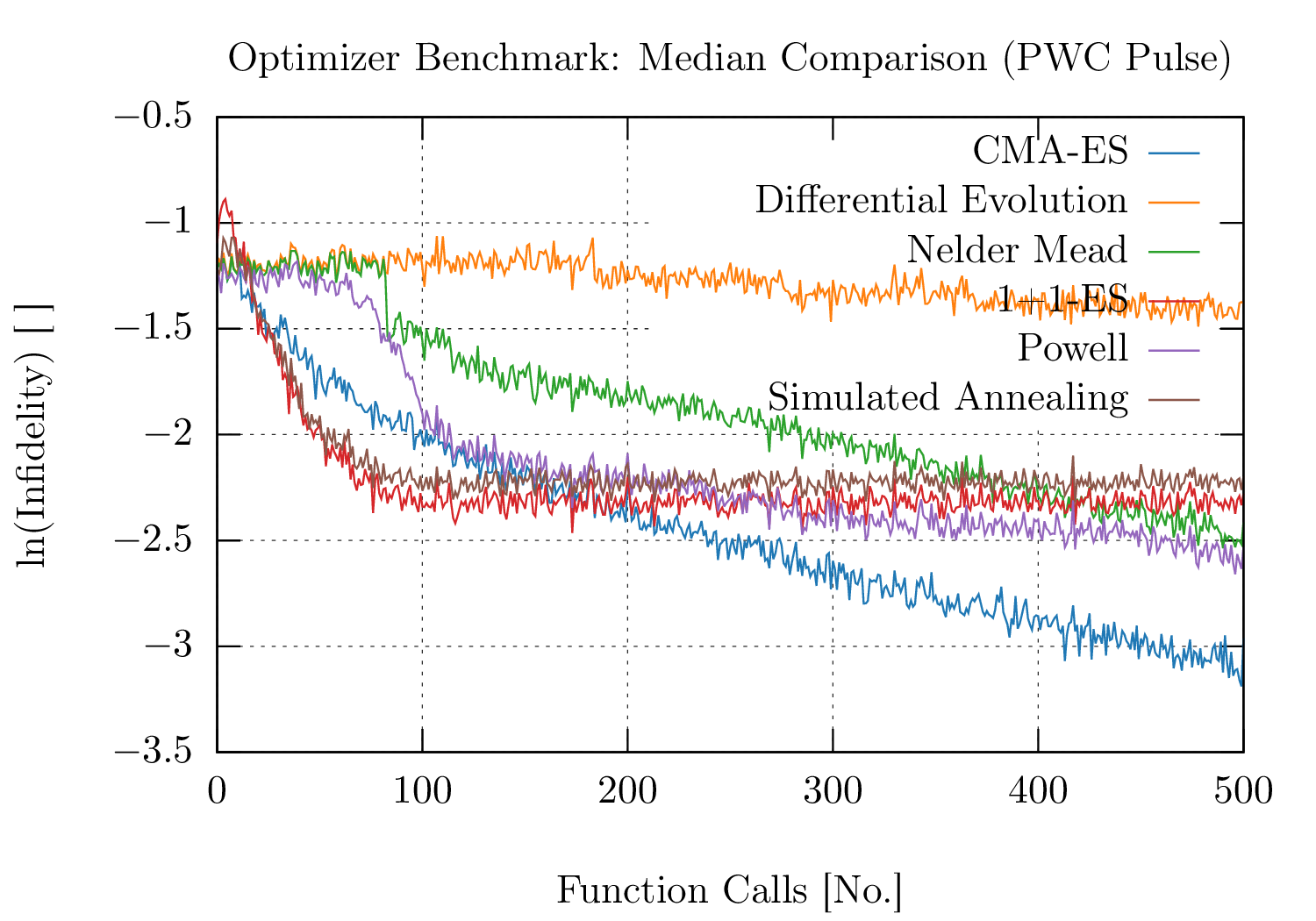}
        \label{fig:pwc_median_comp}
    }

    \caption[Init]{Benchmarking results for the DRAG and PWC pulse simulations. 
    The x-axis shows the number of function evaluations, and the y-axis displays the corresponding 
    loss function value for each parameter setting. 
    Colored lines represent the performance of different optimization algorithms. 
    Figures~\subref{fig:drag_mean_comp} and~\subref{fig:pwc_mean_comp} show the mean values across 
    120 simulations for each optimizer. 
    Figures~\subref{fig:drag_median_comp} and~\subref{fig:pwc_median_comp} show the 
    corresponding medians. 
    CMA-ES achieves the lowest infidelity in all cases.}
    \label{fig:combined_comp}
\end{figure*}

Figures~\ref{fig:drag_mean_comp} and~\ref{fig:drag_median_comp} illustrate the 
behavior of the algorithms during the calibration of the DRAG pulse. All optimizers 
begin with a \(5\%\) detuning from the previously estimated optimal settings, 
simulating an experimental scenario in which only a minimal set of preliminary 
calibrations has been performed to obtain rough parameter estimates. Most algorithms 
converge reliably on the corresponding optimization landscape.

Differential Evolution stands out as an outlier; although it shows some initial 
improvement, it becomes stuck and fails to converge beyond a certain infidelity 
threshold. The Nelder–Mead, \(1+1\)-ES, and Simulated Annealing algorithms 
perform similarly well, achieving rapid improvements within the first 100 
function evaluations but then plateauing at a lower infidelity level. Only 
CMA-ES continues to improve beyond this plateau, albeit initially requiring 
more function evaluations to reach the infidelity level where Nelder–Mead 
and \(1+1\)-ES stagnate. This behavior is even more evident when considering 
the median performance.

Powell performs better than CMA-ES in the early stages but falls behind 
Nelder–Mead, \(1+1\)-ES, and Simulated Annealing, eventually plateauing 
at similar infidelity values. When comparing the median results to the mean, 
notable differences in algorithm behavior become apparent.

While all algorithms show initial improvement, all optimizers except 
CMA-ES tend to stagnate as the number of function evaluations increases. 
Simulated Annealing reaches a lower overall infidelity compared to 
Differential Evolution, Nelder–Mead, and \(1+1\)-ES, but it is 
ultimately outperformed by CMA-ES at higher evaluation counts. 
Powell's method appears to plateau early but exhibits some renewed 
movement in later stages; however, its overall performance remains 
significantly below that of CMA-ES.

The discrepancy between the mean and median performance suggests 
the presence of a few exceptionally well-performing runs for each 
optimizer, which lower the mean but do not affect the median. This 
indicates that while some individual optimization runs succeed, a 
large number of runs struggle to converge, preventing the median 
from reaching lower infidelity levels. CMA-ES is the only algorithm 
that consistently converges in the case of DRAG pulse optimization. 
A likely explanation for this behavior is that the other algorithms 
frequently become trapped in local minima.

Figures~\ref{fig:pwc_mean_comp} and~\ref{fig:pwc_median_comp} illustrate 
the performance of the algorithms in a high-dimensional setting during 
the calibration of the PWC pulse. Differential Evolution shows only slight 
improvement relative to its initial values and generally performs worse 
than the other algorithms. 

While Nelder–Mead previously performed on par with \(1+1\)-ES and 
Simulated Annealing in the low-dimensional setting, it now exhibits a 
significantly longer convergence time, although it ultimately reaches 
a similar final fidelity.

The Powell algorithm converges more effectively than Nelder–Mead 
in this setting but requires more function evaluations than 
\(1+1\)-ES, Simulated Annealing, and CMA-ES. The performance of 
CMA-ES is again noteworthy, achieving the lowest final infidelity 
while requiring fewer iterations than all other algorithms except 
\(1+1\)-ES. The median comparison yields results that are nearly 
identical to those observed in the mean, indicating consistent 
behavior across runs.

Unlike the DRAG pulse case, a large number of optimization runs 
successfully converge in the high-dimensional setting. Even the 
simpler algorithms perform well, which may suggest that the optimization 
landscape contains fewer local minima or presents fewer challenges to 
the optimizers than that of the DRAG pulse. This observation could further 
support the use of more complex pulse shapes in future calibration protocols.

\section{Discussion}\label{sec:discussion}
Given the observed performance of the set of optimizers, it is clear that 
different approaches are better suited for the task of tune-up optimization. 
Although simpler algorithms like Nelder–Mead, Simulated Annealing, and in particular 
\(1+1\)-ES converge with fewer function calls, CMA-ES ultimately reaches a 
higher fidelity in the end. 

Since system calibration is a time-consuming task, algorithms that 
converge quickly remain favorable. The possibility of combining certain 
algorithms to take advantage of both faster convergence and higher accuracy 
is appealing. For example, this might be achieved by starting an optimization 
with Nelder–Mead and switching at a later stage to CMA-ES. However, while 
such an idea seems straightforward at first, its execution may be more 
complex than originally anticipated.

Problems may arise regarding how and when to switch between different optimizers. 
Naively, one might suggest simply switching optimizers after a fixed number of 
function calls. However, this information is only available after prior optimizations 
with the optimizer in question have been completed successfully, and is also 
dependent on the specific system being fine-tuned. 

Another approach would be to switch optimizers once no further improvement in 
fidelity is observed. However, this can be difficult to determine if the loss 
function is noisy and the magnitude of the noise is unknown. Finally, the 
transition between optimizers itself may be non-trivial, as it requires suitable 
initial hyperparameter values for the subsequent optimizer.

Ideally this information is extracted by converting the information the previous 
algorithm has accumulated into suitable parameter values for the next optimizer.  
However, this can be challenging, as parameters of different algorithms maybe 
fundamentally unrelated. 

An example of such a mismatch is the current value of the sampling distribution's standard 
deviation in \(1+1\)-ES. As the only hyperparameter of \(1+1\)-ES, it provides no 
information about the mutation probability required by Differential Evolution. This 
illustrates the difficulty of transferring parameter information between fundamentally 
different optimization strategies.

Another factor that adds complexity to the selection of algorithms for the 
calibration process is the number and nature of their hyperparameters. 
Differential Evolution, for example, has the highest number of hyperparameters 
among all examined algorithms, yet still performs worse than several others.

In contrast, the Powell algorithm stands out for performing well despite having 
no tunable hyperparameters. It should therefore be considered an excellent 
choice for systems where little or no prior knowledge is available.

Additionally, as more complex pulses appear beneficial for increasing the fidelity 
of operations on quantum devices, the observations made in the high-dimensional 
setting are particularly relevant. This setting reveals that even in high-dimensional 
regimes, simple optimizers can perform adequately for the task of calibration.

As such, the choice of optimizer may be of secondary concern, while the choice of 
loss function has a more direct impact on the performance of the optimization 
and thus the achievable fidelity. 
% This is illustrated in 
% Figure~\Kevin{Add landscape plot of ORBIT with target 0 and target 1}, which 
% shows the loss landscapes of two ORBIT functions differing only in the target 
% state. In the left panel, the target state is the ground state, while in the 
% right panel, the target is the first excited state.

% Both landscapes show the characteristic DRAG minimum, but in the plot of 
% the ORBIT sequences targeting the ground state, a valley of decreasing 
% infidelity can be observed toward an amplitude value of zero. This is 
% intuitive, as a zero-amplitude pulse ultimately leaves the system in the ground state. 

% This additional feature of the optimization landscape can pose 
% challenges for optimizers, which may become trapped in such 
% regions and produce suboptimal results.

Finally, while it was previously noted that the overall choice of 
algorithm may not be critical for achieving convergence, it is 
still worth emphasizing that the final infidelity achieved by CMA-ES is lower in 
every setting compared to all other algorithms. As such, CMA-ES appears to be the 
ideal choice for fine-tuning pulses when the final fidelity of the operation 
is the most important performance metric.

\section{Conclusions}\label{sec:conclusions}
To summarize our findings regarding the choice of optimization algorithm, we 
recommend the use of CMA-ES from the set of optimizers investigated in 
this work. CMA-ES performs well across a range of settings, exhibits 
strong robustness to noise, effectively handles local extrema, and 
requires relatively few function evaluations. With its low number of 
initial hyperparameters and its ability to adapt them during optimization, 
CMA-ES is also relatively easy to set up—especially when calibrating 
systems with little prior information. Similar conclusions have been 
drawn independently in~\cite{rapin2019}.

Notably, CMA-ES consistently achieved the highest fidelities, highlighting 
that the choice of optimization algorithm can have a direct impact 
on calibration outcomes. This naturally raises the question of whether 
even more effective algorithms—potentially tailored specifically for 
quantum system calibration—could lead to further improvements in achievable fidelity.

Another key takeaway from this investigation is the critical importance of 
the loss function. Over the course of this project, it became clear that 
the nature of the loss function plays an even more decisive role in the 
success of automated tune-up than the choice of optimizer itself. We 
therefore see a strong need for future research into how different 
figures of merit perform as loss functions in the context of automated calibration.

Finally, we note that our work resonates with broader efforts in the computer science 
community. For example, frameworks such as \textit{Nevergrad}, developed by Facebook 
Research, provide a collection of gradient-free optimizers; the \textit{COCO} platform 
(COmparing Continuous Optimisers) underpins the well-established 
Black-Box-Optimization-Benchmarking (BBOB) workshops; and similar conclusions 
regarding the effectiveness of CMA-ES have been drawn by J. Rapin et al.~\cite{rapin2019}. 

\section{Acknowledgement}
This work was supported by the German Federal Ministry of Education and Research (BMBF) as part 
of the Digital Analog Quantum Computer (DAQC) initiative within the framework 
"Quantenprozessoren und Technologien für Quantencomputer." The 
authors gratefully acknowledge this support, which made the research presented in this paper possible.

\bibliography{bibliography}  % Ensure your .bib file is named 'bibliography.bib'

% \section{Appendix}

% \subsection{Similar Efforts in Computer Science}
% \begin{itemize}
%     \item Nevergrad \newline
% A collection of gradient-free optimizer algorithms. Written and maintained by
% Facebook Research
%     \item COCO (COmparing Continuous Optimisers), \newline
% A platform for the comparison of optimizers and linked to the ongoing series of
% Black-Box-Optimization-Benchmarking (BBOB) workshops.
%     \item Similar conclusions reached by J. Rapin et al. Proceedings of the Genetic and
% Evolutionary Computation Conference Companion (2019)
% \end{itemize}

% \Kevin{add maybe benchmarks of optimizers on analytical functions here? + analytical landscapes
% if not: remove mention of analytical benchmarks from abstract}
% \subsection{Technical Details of Implementation}

% \subsubsection{Transformation to Optimizer Space}
% % In order to achieve a better convergence we transform the goal function landscape by applying the following function 
% % \begin{equation}
% %     F_x(x, y, s) = h(x, s) * y + g(x, s)
% % \end{equation}
% % with $y = f(x)$
% % \begin{equation}
% %     h(x,s) =     
% % \end{equation}
% % \begin{equation}
% %     g(x,s) =     
% % \end{equation}

\end{document}